# Is there a relationship between wettability and rates of equilibration of the H-bonded oligomer PMMS under confinement?

Sara Zimny[1], Magdalena Tarnacka[2], Monika Geppert-Rybczyńska[1], Kamil Kamiński[2*]

[1] *Institute of Chemistry, University of Silesia in Katowice, Szkolna 9, 40-006 Katowice, Poland*

[2] *Institute of Physics, University of Silesia in Katowice, 75 Pułku Piechoty 1, 41-500 Chorzów, Poland*

*Corresponding author:* kamil.kaminski@us.edu.pl

**Abstract**

In this paper, we investigated the annealing experiments of poly(mercaptopropylmethylsiloxane, PMMS) confined within two types of porous templates (anodic aluminium oxide, AAO, and silica) characterized by different pore diameter, $d = $ 8-120nm, using different thermal protocols (varying significantly in cooling/heating rate) by means of Broadband Dielectric Spectroscopy (BDS) supported by the complementary Differential Scanning Calorimetry (DSC) and temperature-dependent contact angle, $\theta$, measurements. It was found that relaxation times obtained from routine temperature-dependent dielectric investigations deviate from the bulk behavior when approaching the glass transition temperature. Importantly, this confinement induced effect can be easily removed by the annealing experiments performed at some specific range of temperatures. The analysis of the dielectric data collected during isothermal experiments of confined samples that was beforehand cooled with different rates revealed that (i) constant rates of annealing gets longer with cooling and weakly depend on the rate of cooling, and (ii) activation energy of the equilibration process, $E_a$, varies with the reduction of the pore diameter and material the porous template is made of. In fact, there is significant reduction in $E_a$ from ~62 to ~23 kJ/mol obtained for the annealing process carried out in AAO ($d = $ 10 nm) and silica ($d = $ 8 nm) membranes, respectively. Such significant change in $E_a$ can be explained taking into account temperature-dependence of $\theta$ of PMMS indicating a notable change in wettability between both surfaces upon cooling. As a consequence, one can expect that the mass exchange between interfacial and core molecules as well as adsorption-desorption processes occurring at the interface at lower temperatures must be affected. Our data provide a better understanding on the annealing processes occurring in the liquids confined in mesopores.

## I. Introduction

Soft materials subjected to nanospatial restriction has already been studied for decades. Over this time, it was shown that their basic physical parameters, such as viscosity, diffusion, relaxation times, gas permeability, creep behaviour, as well as phase and glass transition temperature can be altered/tuned by varying the kind of confinement (hard vs. soft), degree of the confinements (one-, two- or three-dimensional), finite size as well as change in the molecular interactions between host and guest molecules.[1–3] The last can be achieved by proper physical or chemical modification of the substrate to make it either more attractive, repulsive, rougher, hydrophilic *etc*. However, further intensive studies on the polymers supported as layers of varying thicknesses clearly indicated that all confinement induced variation in the above-mentioned parameters can be erased fully or partially by the annealing process.[4–7] Just to mention, experiments on polystyrene, polyvinyl acetate or poly(methyl methacrylate),[8] where after prolonged annealing (time scale of experiments significantly exceeded relaxation time of segmental process) the recovery of bulk-like rheological properties (and/or segmental relaxation times of the spin casted macromolecules) was reached. Importantly, this process was found to depend on the polymer molecular weight, $M_w$.[9] Thus, at the same time of annealing, polymers of lower $M_w$ restored features of the macroscale systems; while, those characterized by much higher molecular weight did not fully equilibrate even after one week. This phenomenon was assigned to a removal of mechanical stress generated during the preparation of the thin films.[9] Further investigations indicated that restoring bulk-like dynamics of confined polymers can be related to the equilibration of the free volume that propagate through the entire film as well as a formation of the irreversibly adsorbed layer.[9–11]

Nevertheless, as far as there are numerous experimental and simulation studies on the behaviour of polymers deposited as thin films during annealing, much less is done in case of the materials infiltrated into mesoporous matrices. From one hand, it is justified since in such systems, the guest molecules are being in contact with strongly curved surface and it is very difficult to probe directly properties of such systems. However, on the other hand, the effects observed for the polymer thin films should be magnified in pores geometries due to significantly higher surface-to-volume ratio for this kind of confinement.[12] In fact, it was revealed that similarly to the thin films, one can detect and monitor the equilibration process, manifested as the shift of the segmental relaxation process towards lower frequencies upon isothermal time-dependent measurements for the polymers infiltrated into mesoporous matrices.[1,13–15] Interestingly, although some coupling between global dynamics and rate constant of sample equilibration was detected, the time scale of this process significantly exceeded segmental and global mobility of the chains.[1] In contrast to the data published for poly(propylene glycol, PPG) infiltrated into anodic alumina porous (AAO) templates, it was possible to recover the bulk-like segmental relaxation times of poly(phenylmethylsiloxane, PMPS).[4] However, the shape of the segmental relaxation of the bulk sample was never reproduced.[16] Additionally, it should be highlighted that similarly to the data obtained for the polymer thin films, the time-scale of the bulk-like recovery of the dynamics depends on the molecular weight of the polymer.[17] These experiments served as a foundation to formulate a hypothesis that the equilibration of the polymers infiltrated into porous membranes is due to extremely slow viscous flow.[4] The same conclusion was derived by some of us for the low-molecular-weight polar system infiltrated into mesoporous silica and alumina pores.[18] Furthermore, one can recall the recent work by Kardasis *et al.*,[15] who performed dielectric studies on the both linear and star-shaped poly(*cis*-1,4-polyisoprene, PI) confined in alumina mesoporous membranes using two different thermal protocols. Importantly, in agreement with the previous reports, they showed that there is a critical temperature below, which the confined sample enters non-equilibrium region mimicking liquid-glass transition. It was also demonstrated that this critical temperature depends on the pore diameter. Moreover, the equilibration times were strongly dependent on the properties of examined macromolecules (especially their molecular weight and architecture). As shown, the star-shaped polymers required more time to the recovery of bulk-like dynamics with respect to the linear ones.[15] In this context, one can also mention data obtained for the low-molecular-weight samples, which undergo the equilibration above some specific (critical) temperature assigned, in their cases, as the vitrification of the interfacial layer.[5,6,19]

In this paper, we explore further the idea previously reported by Kardasis *et al.*[15] and perform a series of the annealing experiments on poly(mercaptopropylmethylsiloxane, PMMS) confined within two types of porous templates (anodic aluminium oxide, AAO, and silica) characterized by different pore diameter, $d = 8$-$120$ nm, using different thermal protocols (varying significantly in cooling/heating rate). For those systems, we carried out a series of annealing experiments and estimated the activation barrier for the observed equilibration and its dependency on the cooling/heating rates, pore diameter, and wettability.

## II. Materials and Methods

**Material.** Poly(mercaptopropyl)methylsiloxane homopolymer (labeled as PMMS, and characterized by the molecular weight of $M_n = 2\,400$ g/mol, determined by Viscotek TDA 305 triple detection calibrated with a narrow polystyrene standard, THF as eluent, for details please see Ref. [20]) was purchased from Gelest and used as received. However, for the contact angle measurements it was dried for 48 h at temperature $T = 303$ K under pressure $p = 10$ mbars. The mesoporous anodic aluminium oxide, AAO, membranes used in this study were supplied from InRedox; whereas those made of silica were produced by us according to the procedure described in Ref. [21,22]. Both types of applied templates are composed of uniaxial channels (open from both sides) with well-defined pore diameters. The chemical structure of examined compound is presented in **Figure 1(a)**.

Before filling, all types of membranes were dried in an oven at $T = 423$ K under vacuum to remove any volatile impurities from the nanochannels. After cooling, they were used as a constrain medium. For that purpose, the templates were placed in a small glass flask containing PMMS. The whole system was maintained at $T = 298$ K in a vacuum ($10^{-2}$ bar) for $t = 1$ h to let the samples flow into the nanocavities. Samples were finally annealed at $T = 353$ K under a vacuum ($10^{-2}$ bar) for $t = 1$ h and weighed thereafter. The complete filling was obtained by a series of repeated infiltration procedure until the weight of the templates before and after was constant. After filling, the excess sample on the surface of the membranes was removed with a paper towel. The filling degree reaches ~90%. Those values are calculated considering the porosity of membranes and the assumption that the density of the infiltrated material does not change along the pore radius and that the shape of the pore is cylindrical. The filling degree value depends on the porosity of the applied membrane and was different for alumina and silica membranes.

**Broadband Dielectric Spectroscopy (BDS).** BDS measurements were carried out on heating after a fast quenching of the liquid state in a wide range of temperatures (175 – 243 K) and frequencies ($10^{-1} - 10^6$ Hz) using a Novocontrol spectrometer, equipped with Alpha Impedance Analyzer with an active sample cell and Quatro Cryosystem. Dielectric measurements of bulk samples were performed in a parallel-plate cell (diameter: 15 mm, gap: 0.1 mm) as described in Ref. [20]. AAO membranes filled with studied alcohols were also placed in a similar capacitor (diameter: 10 mm, membrane thickness: 0.005 mm). Nevertheless, the confined samples are a heterogeneous dielectric consisting of a matrix and an investigated compound. Because the applied electric field is parallel to the long pore axes, the equivalent circuit consists of two capacitors in parallel composed of $\varepsilon^*_{compound}$ and $\varepsilon^*_{AAO}$. Thus, the measured total impedance is related to the individual values through $1/Z_c^* = 1/Z_{compound}^* + 1/Z_{AAO}^*$, where the contribution of the matrix is marginal. The measured dielectric spectra were corrected according to the method presented in Ref. [8].

The obtained dielectric loss spectra were analyzed by the superposition of two Havriliak-Negami (HN) functions with an additional term related to the dc conductivity:[23]

$$\varepsilon(\omega)" = \frac{\sigma_{dc}}{\varepsilon_0 \omega} + Im \sum_{i=1}^{2} \left( \varepsilon_\infty + \frac{\Delta \varepsilon_i}{[1+(i\omega\tau_i)^{\alpha_i}]^{\beta_i}} \right) \qquad (1)$$

where $\alpha$ and $\beta$ are the shape parameters related to the symmetric and asymmetric broadening of relaxation peaks, $\Delta\varepsilon$ is related to the dielectric strength, $\tau_{HN}$ is the HN relaxation time, $\varepsilon_0$ is the vacuum permittivity and $\omega$ is an angular frequency, where $\omega = 2\pi f$.

**Differential Scanning Calorimetry (DSC).** Calorimetric measurements were carried out by Mettler-Toledo DSC apparatus equipped with a liquid nitrogen cooling accessory and an HSS8 ceramic sensor (heat flux sensor with 120 thermocouples). Temperature and enthalpy calibrations were performed by using indium and zinc standards. The sample was prepared in an open aluminum crucible (40 μL) outside the DSC apparatus. Samples were scanned at various temperatures at a constant heating rate of 5, 10, and 20 K/min.

**Contact Angle Measurements.** The contact angle, $\theta$, was measured at nine temperatures in the wide temperature range, $T = 263 - 298$ K, by the DSA 100S Krüss Tensiometer, GmbH Germany with the Advance software. Thanks to the measuring chamber adopted to work in the wide temperature range with a humidity control, it was possible to measure contact angle below the freezing point of water. The general description of procedures has been presented previously.[24] The contact angle measurements on the solid surfaces have been repeated dozen or more times. The temperature measurements uncertainty was $\pm$ 0.1 K. The precision of contact angle measurements was 0.01°, and the estimated uncertainty was not wore than $\pm$ 2°.

The surface energy for native silica was 67.6 mJ/m$^2$ with dominant non-dispersive part equal 66.6 mJ/m$^2$. For alumina respective value was 59.0 mJ/m$^2$ with non-dispersive part 55.6 mJ/m$^2$.[25]

## III. Results and Discussion

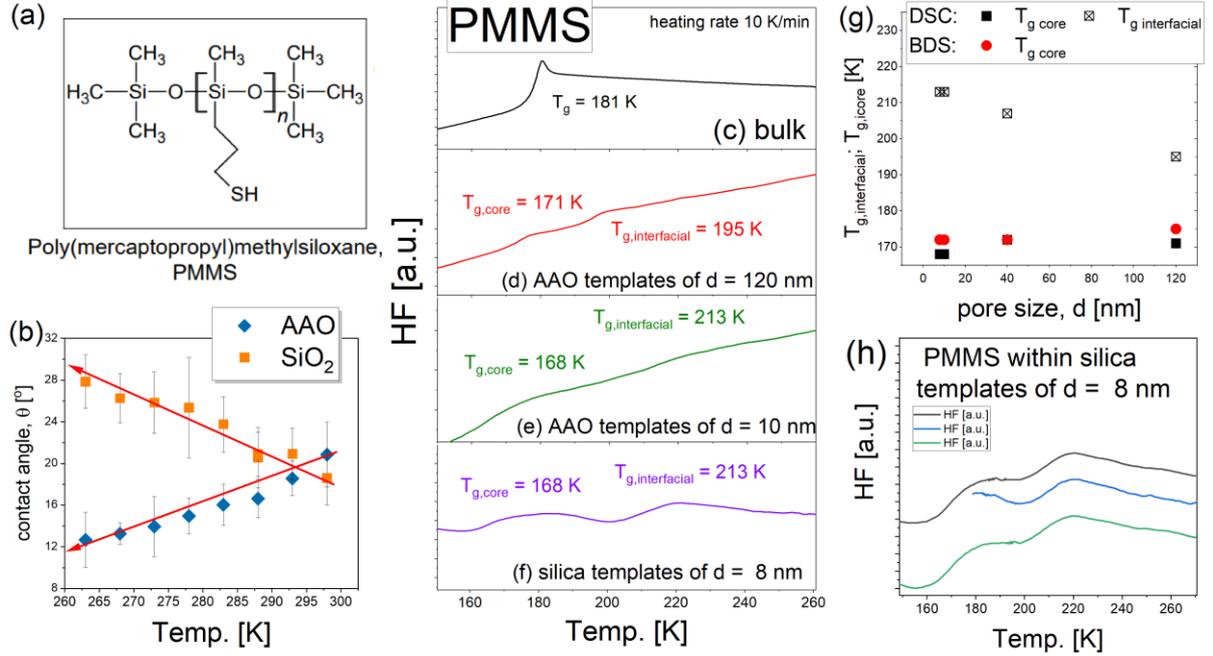

**Figure 1**. (a) The chemical structure of examined PMMS; (b) Temperature dependences of contact angle, $\theta$, of PMMS measured on two different surfaces, alumina (blue points) and silica (orange points). (c-f) DSC thermograms obtained for the bulk sample (c) and PMMS infiltrated within AAO (d,e) and silica (f) templates of different pore size, $d$; (g) the pore diameter dependence of the glass transition temperatures, $T_g$, obtained from both calorimetric and dielectric measurements. Data taken from **Table 1**. (h) the comparison of DSC curves obtained in different temperature ranges.

**Table 1**. The glass transition temperatures of core, $T_{g,core}$, and interfacial, $T_{g,interfacial}$, molecules obtained from both calorimetric and dielectric measurements.

| sample | DSC measurements | | BDS measurements |
|---|---|---|---|
| | $T_{g,core}$ [K] | $T_{g,interfacial}$ [K] | $T_{g,core}$ [K] |
| bulk | 181 | --- | 172 |
| AAO templates | | | |
| $d = 120$ nm | 171 | 195 | 175 |
| $d = 40$ nm | 172 | 207 | 172 |
| $d = 10$ nm | 168 | 213 | 172 |
| Silica membranes | | | |
| $d = 8$ nm | 168 | 213 | 172 |

As a first step of our investigation, we performed the calorimetric measurements of all examined systems. The representative DSC thermograms obtained during heating with 10

K/min are shown in **Figures 1(c-f)**. As shown, the DSC curve of bulk PMMS display one endothermic process related to the glass transition located approximately at $T_g = 181$ K, whereas the porous materials exhibit two endothermic processes, which is a manifestation of the so-called double glass transitions (DGT)[26] occurring below and above the bulk $T_g$. It should be mentioned that DGT phenomenon is a common features of various low- and high-molecular-weight glass formers incorporated into porous templates.[26–30] It results from the distribution of interactions induced by the applied constrain medium. Thus, two fractions of molecules of different dynamics can be distinguished: (1) "interfacial" layer associated with the molecules placed in the close vicinity of the pore walls characterized by limited mobility, and $T_g$ higher than the bulk system ($T_{g,interfacial}$) and (2) "core" molecules located in significant distance from the pore walls (closer to the center of the porous channels) characterized by the $T_g$ lower than the bulk ($T_{g,core}$). Values of $T_g$s are listed in **Table 1** and in **Figure 1(g).** At this point, one can mention that some approaches imply that the presence of DGT (and especially $T_{g,interfacial}$) is conditional and can be observed after sample reaches the spinodal temperature (which is close to ~$T_{g,core}$).[31,32] Nevertheless, so far, this effect seems to be limited to few polymeric materials,[31] as so far for the low-molecular-weight glass formers infiltrated within AAO templates, it was not observed.[18] Thus, herein, we also verified the correlation between the DGT and the spinodal temperature. For this purpose, we performed three cooling scans: (1) to 160 K ($T < T_{g,core}$), (2) to 190 K ($T_{g,core} < T < T_{g,interfacial}$), and (3) again to 184 K ($T < T_{g,core}$). Note that each cooling scan was followed by heating to the room temperature with the heating/cooling rate of 10 K/min. Representative DSC thermograms obtained in the above-mentioned manner for PMMS infiltrated into silica templates of $d = 8$ nm are presented in **Figure 2(h)**. Interestingly, for all recorded curves, a well-visible $T_{g,interfacial}$ can be seen. Note that the same results were obtained also for PMMS incorporated within AAO templates of both $d = 10$ nm and $d = 120$ nm, see **Figure S1(a)**. This indicates: (1) no existence of the spinodal temperature in the case of examined compound and (2) that DGT is a coherent feature of confined PMMS.

Moreover, as illustrated in **Figure 1(g)**, the difference between both $T_g$s ($T_{g,interfacial}$ and $T_{g,core}$) increases with a decreasing pore diameter.[26–30] Interestingly, there is no difference between the behavior of PMMS infiltrated in either AAO templates of $d = 10$ nm or silica membranes of $d = 8$ nm, see **Table 1**. Although these two types of porous templates have a comparable pore diameter, they are characterized by often significantly different wettability. Note that recent studied on various confined glass formers showed that both surface, alumina and silica, differs in contact angle, $\theta$,[18,19,25] which provides a measure of wettability; thus, the better spearing out of a liquid drop on the examined surface, the smaller $\theta$, i.e., for water, $\theta < 90º$ indicates hydrophilic surface. In this context, to quantify wettability of PMMS on both surfaces, we carried out additional measurements of $\theta$ as a function of temperature. One can recall that up to data, all the discussion on the wettability of materials characterized by relatively low glass transition temperatures were made based on measurements of $\theta$ at room temperature. Nevertheless, herein, we are able to see how the contact angle changes with temperature. Values of measured contact angles are shown in **Figure 1(b).** Interestingly, at $T = 295$ K, PMMS on both surfaces is characterized by comparable contact angle, $\theta \sim 20º$. However, one can see that the examined compound behaves differently upon cooling; while for alumina surface, the value of $\theta$ decreases ($\sim 20º$) with lowering temperature (blue points in **Figure 1(b)**) an opposite trend can be observed for the silica substrate (orang points in **Figure 1(b)**). Consequently, at $T = 263$ K, PMMS is characterized by $\theta \sim 12º$ and $\theta \sim 28º$ at alumina and silica surface, respectively. Nevertheless, although we observed a different wettability of PMMS on both flat surfaces at low temperature no difference in $T_g$s for PMMS infiltrated into both mesoporous membranes was detected (see **Table 1**).

In next step, we carried out the additional series of calorimetric measurements, upon which, both materials (PMMS incorporated within both AAO templates of $d = 10$ nm and silica membranes of $d = 8$nm) were annealed at $T_{g,bulk} + 50$ K ($T_{anneal} \sim 240$ K) for the same amount of time, $t = 240$ min. Surprisingly, the performed annealing in no way affect the examined samples, as both $T_g$s were the same as before the annealing, independently to the applied porous materials, see **Figure S1(b)**. This might imply that although the examined surfaces differ in wettability, the difference between them might not be so significant to govern the overall behavior of PMMS molecules adsorbed or being in the close vicinity to the pore walls.

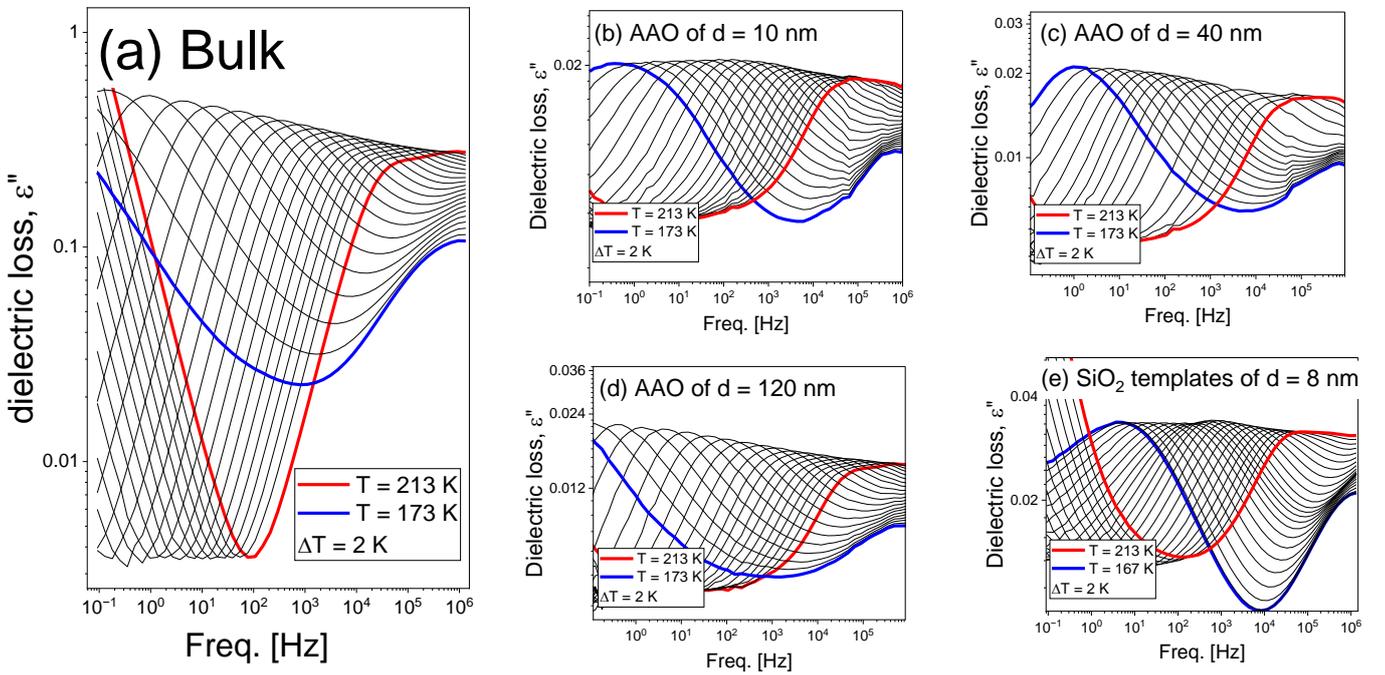

**Figure 2**. (a-d) Dielectric loss spectra obtained for the bulk sample (a) and PMMS infiltrated within AAO (b,c,d) and silica (e) templates of different pore size, $d$.

In **Figure 2**, we presented dielectric loss spectra measured for the bulk sample (**Figure 2(a)**) and PMMS infiltrated into AAO and silica membranes of given pore diameter (**Figure 2(b-e)**). As shown in **Figure 2(a)**, one can easily identify dc conductivity and main relaxation process which at higher temperature is bimodal. Upon cooling, it becomes narrower and in the vicinity of $T_g$, it is visible as a single relaxation process. As discussed in our previous papers,[20,33] the characteristic shape of this mode is related to the contribution of the sub-Rouse (appearing at lower frequencies) and segmental relaxation process (detected at higher frequencies) of similar time scale to the dielectric response.[20,33] Importantly, the same scenario is also noted for the confined samples. To illustrate this finding in a better way, a loss spectra obtained at higher temperatures, characterized by the similar segmental relaxation time were superposed to coincide at maximum in **Figure S2**. It can be easily found that apart of the common broadening of the dielectric response for the samples infiltrated into porous media a clear bimodal character of the main process is preserved. We also detected some small variation in the amplitude of the sub-Rouse vs. segmental relaxation process for PMMS infiltrated in silica and alumina pores. However, this effect is not significant.

Having this in mind, we decided to analyse the data presented in **Figure 2** with the use of either single HN or two HN functions with the conductivity term (Eq. (1) in **Experimental section**) for the samples measured respectively at higher and lower temperatures. As discussed above, this pattern of data analysis was related to the systematic narrowing of the main relaxation process and, the fact that, losses its bimodal character in the vicinity of $T_g$. Representative dielectric data along with the respective fits are presented in **Figure S3**.

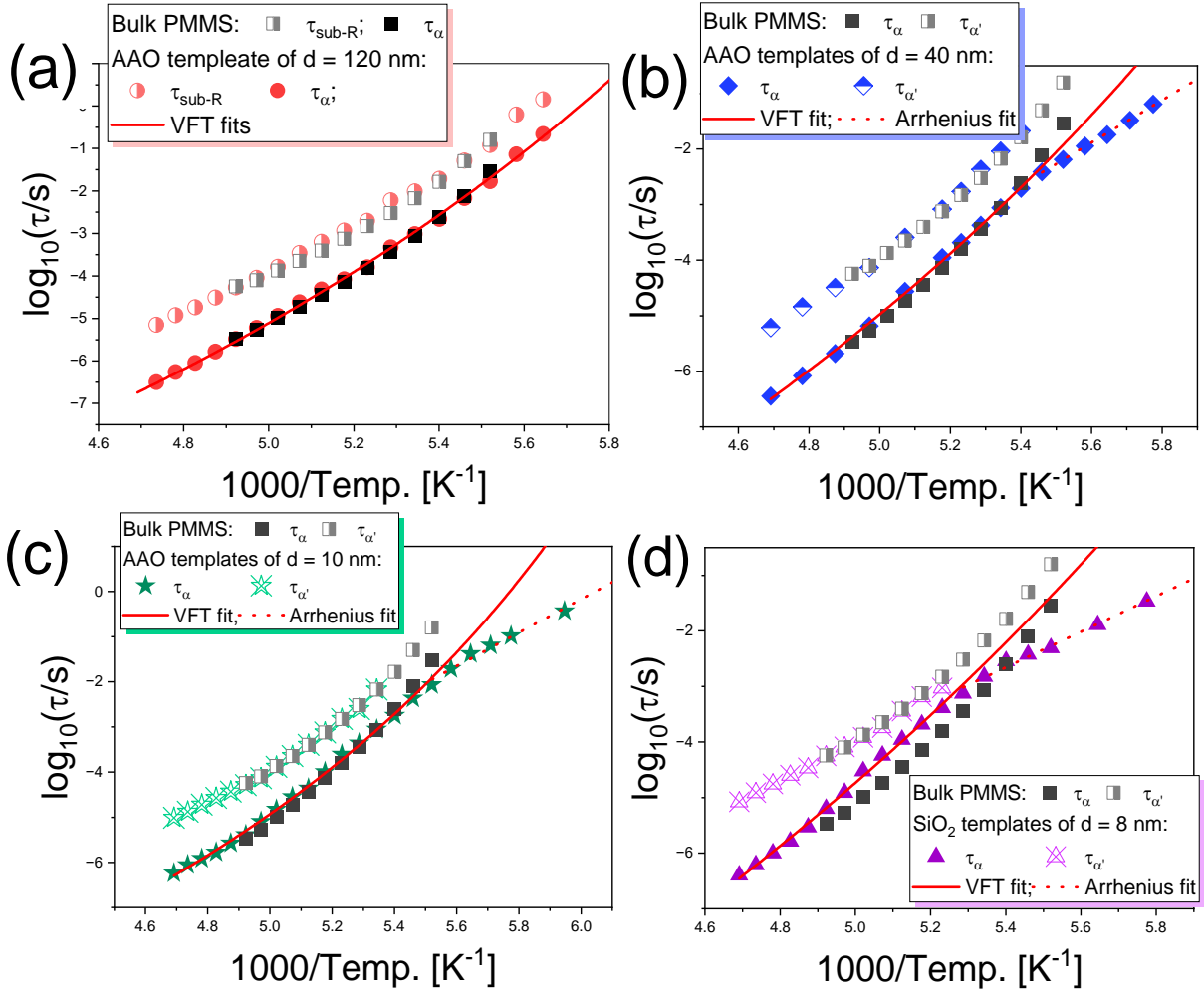

**Figure 3**. Temperature dependences of sub-Rouse, $\tau_{sub-R}$, and segmental, $\tau_\alpha$, relaxation times obtained from the analysis of the data presented in **Figure 2** with the use of the two HN functions with the conductivity term (eq. (1)).

After analysis of the data was completed segmental and sub-Rouse mode relaxation times were plotted vs reciprocal temperature in **Figure 3** for the confined samples along with the ones obtained for the bulk PMMS. As can be seen, both sets of data coincide very well at higher temperatures, where we monitored the bimodal process. At lower temperature, where only single process was observed relaxation times of segmental mode of the confined systems start to deviate from the bulk behaviour. What is more, this notable fluctuation in segmental dynamics shifts to a higher temperature (shorter relaxation times) with systematic decrease in pore diameter and is independent on the porous media (silica or alumina). It is worthwhile to stress that such scenario is commonly reported for the small and high molecular weight samples infiltrated into pores. There are numerous hypotheses trying to explain this universal finding (please see ref.[34–39]). However, the one assuming vitrification of the interfacial (adsorbed) molecules seems to be the most appropriate. Although last papers questioned this interpretation based on the simple core-shell model.[40]

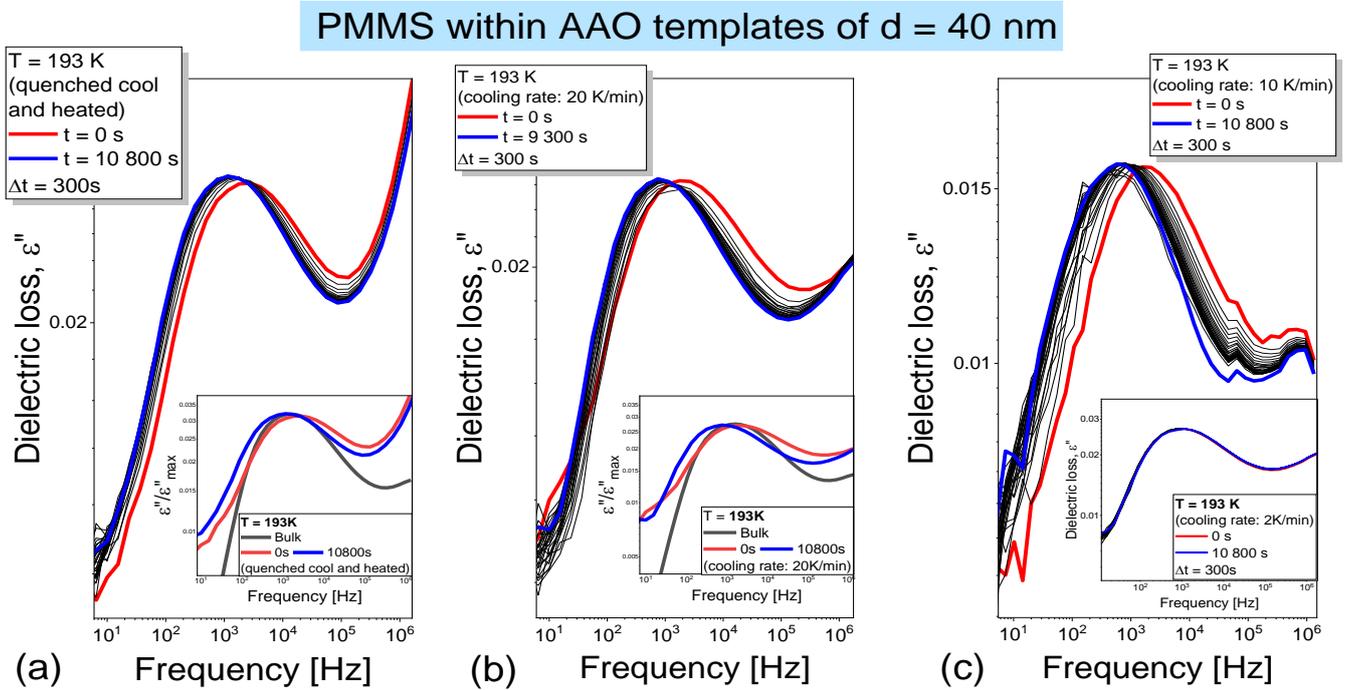

**Figure 4** Time evolution of dielectric loss spectra collected upon the annealing experiments of PMMS within AAO templates of $d = 40$ nm at $T_{anneal} = 193$ K *via* different cooling rates. The insets show the comparison of dielectric loss spectra shapes before and after the annealing in comparison to the bulk material.

Having in mind results of calorimetric and dielectric investigations and inspired by the work by Kardasis *et al.*,[15] we decided to perform systematic studies on the effect of annealing of the confined PMMS below temperature where a deviation in the segmental dynamics from the bulk behaviour was noted. Similarly, to the data reported in Ref. [15], two different protocols were applied. The first one (1) relied on the cooling of the sample with various rates to a given annealing temperature, $T_{anneal}$; while, the second (2) protocol is based on the quenching (~ 40

K/min) of the sample deep below $T_g$ of the bulk material and reheating (with constant heating rate of 10 K/min) to the desired $T_{anneal}$. In both cases after reaching $T_{anneal}$, the isothermal time-dependent measurements were started. In **Figure 4**, we have shown representative dielectric loss spectra collected during the annealing of PMMS within AAO templates of $d = 40$ nm at $T_{anneal} = 193$ K obtained *via* different pathways. It is well visible that for each sample, we observed shift of the segmental process to the lower frequencies with time. However, this effect was less or more pronounced and strongly depended on the cooling rate. It seems to be the least significant for the PMMS cooled with the lowest rate (see the inset in **Figure 4(c)**), respectively. In addition, we also found that (1) the distribution of the relaxation time of the main process does not change with the time of annealing and (2) confined induced fluctuations in segmental dynamics are erased after experiment was completed. The latter finding is well proven by the insets in **Figure 4(a,b),** where we compared the loss spectra of the bulk PMMS and sample infiltrated into alumina pores prior and after equilibration. In each case, the position of the main peak after annealing fairly agrees with the one obtained for the bulk system at given temperature.

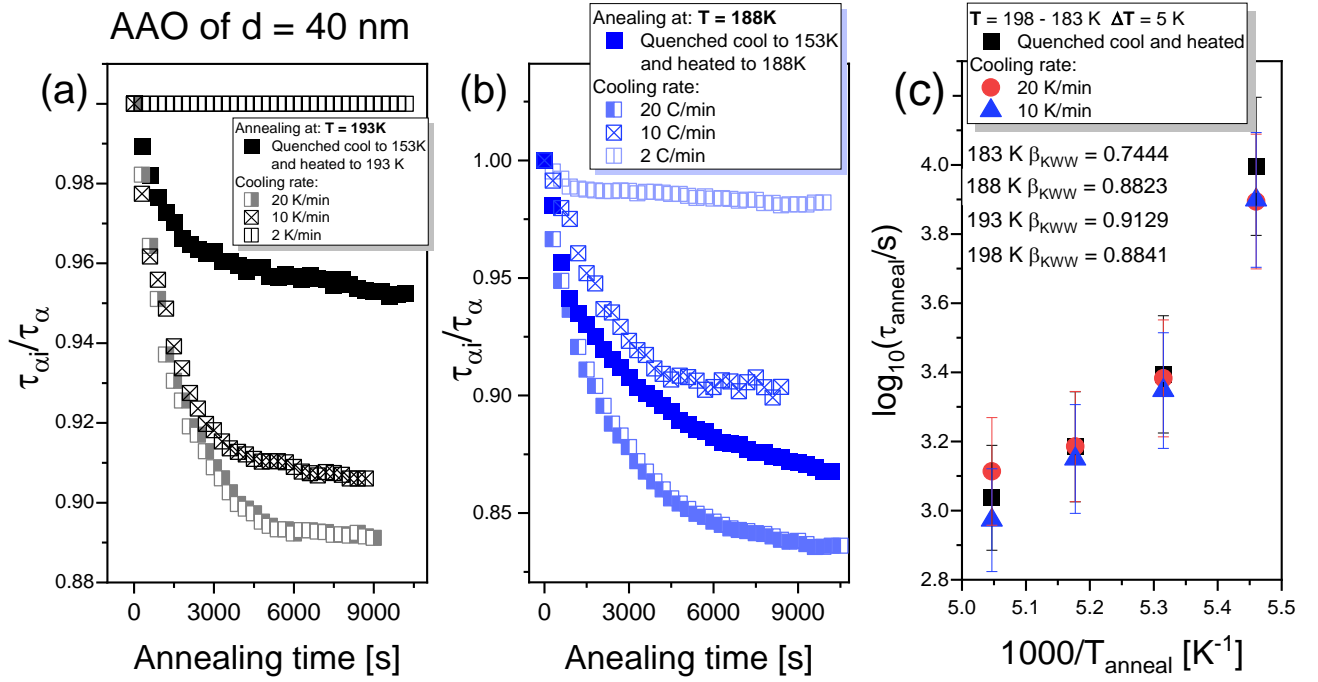

**Figure 5**. (a,b) Time evolution of the renormalized segmental relaxation times from the annealing experiments performed for PMMS within AAO templates of $d = 40$ nm cooled with different rates at two representative $T_{anneal}$; (c) Temperature dependences of $\tau_{anneal}$ obtained from eq. (3).

Taking advantage of the described above dielectric investigations in the next step we decided to plot the time evolution of the segmental relaxation times during annealing at $T_{anneal} = 193$ K for the samples obtained *via* different pathways, **Figure 5(a)**. Note that in order to present all the data in a clear fashion we rescaled them according to the following formula $\tau_i/\tau_\alpha$ where $\tau_i$ and $\tau_\alpha$ stands for the initial segmental relaxation time, segmental relaxation time at given $t$. In order to describe the obtained curves we applied the following equation:

$$\tau = A \exp\left(-\left[\frac{t}{\tau_{anneal}}\right]^\beta\right) + \tau_\infty \qquad (2)$$

where $A$, $\tau_{anneal}$, $\beta$ and $\tau_\infty$ are preexponential factor, characteristic annealing time, stretching exponent, equilibrium relaxation time respectively. This equation is very often used to describe the data obtained during physical aging or annealing of various samples. From the fitting of the time evolution of the renormalized segmental relaxation times to eq. (2), we were able to obtain characteristic $\tau_{anneal}$, which were further plotted vs reciprocal temperature for the samples obtained *via* different pathways. Importantly, just as in the case of Ref. [15], it was found that the characteristic times of annealing are almost the same and do not depend on the cooling rate or thermal history of the sample. Having in mind this important finding we decided to limit the number of further measurements and perform annealing experiments at different temperatures for the samples obtained *via* fast quenching (protocol 2) below $T_g$ followed by the heating to the desired temperature for the PMMS infiltrated into alumina and silica pores having $d = 10$ and $d = 8$ nm, respectively.

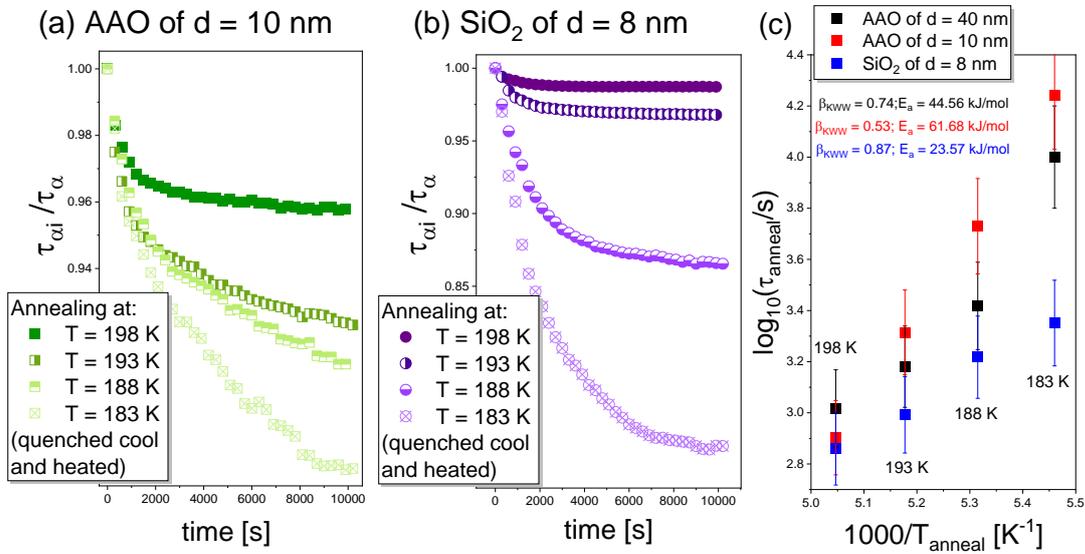

**Figure 6.** (a,b) Time evolution of the renormalized segmental relaxation times for the annealing experiments performed in AAO and silica templates of $d =10$ nm and $d = 8$ nm, respectively; (c) Temperature dependences of $\tau_{anneal}$ obtained *via* eq. (3).

In **Figure 6**, we presented time evolution of the renormalized segmental relaxation times for the annealing experiments performed in silica and alumina pores. From this data it is well-visible that irrespectively to the difference in wettability of PMMS on both surfaces in both cases we observe systematic shift of the segmental process towards lower frequencies with the time at low temperatures. What is more, as far as annealing experiments were performed at higher temperatures the magnitude of the shift of the segmental process was similar. However, at lower temperatures, the effect of annealing was much more pronounced for the sample measured in silica pores. One can suppose that this finding must be related to the adsorption - desorption phenomena occurring at the interface since wettability of PMMS on silica and alumina varies at low temperatures. Nevertheless, in each case, we observed elongation of the process with the lowering annealing temperature. Having determined $\tau_{anneal}$, in next step, we plotted them vs reciprocal temperature in **Figure 6(c)**. As can be observed $\log \tau_{anneal}(1/T)$ follow linear dependency irrespectively of the pore diameter and mesoporous template. Although, they have different slopes suggesting different activation energy for the equilibration process. To determine this important parameter, we applied the Arrhenius equation:

$$\tau = \tau_\infty \exp\left(\frac{E_a}{k_B T}\right) \qquad (3)$$

where $\tau_\infty$ and $E_a$ are preexponential factor and activation barrier, respectively. From the fitting eq. (3) to the data presented in **Figure 6(c),** we obtained that activation energy of the equilibration process depends on the pore diameter as well as matrix. For the equilibration experiments performed in alumina pores, one can see that $E_a$ increases from $E_a \sim$ 44.6-61.7 kJ/mol with a reducing pore diameter. However, unexpectedly, it decreases significantly to 23.6 kJ/mol for the experiments carried out in silica templates (of $d = 8$ nm). Importantly, the estimated herein activation barrier are much smaller with respect to the one obtained for the equilibration of polylactide in alumina pores ($E_a \sim 282 \pm 5$ kJ/mol).[41] Nevertheless, interestingly, all these determined $E_a$ are comparable to the activation barrier for the adsorption process of polystyrene (PS) deposited on silicon oxide (where $E_a \sim 60$ kJ/mol).[42]

Therefore, it seems that the equilibration process we observed during the annealing experiments might be, in fact, related to the adsorption or desorption phenomenon occurring at the interface (including, i.e., local noncooperative rearrangements and mass exchanges processes between interfacial and core molecules) as reported for the thin polymer films.[42] Herein, it should be commented that much different activation energy of the equilibration process in silica and alumina matrix can be related to a different wettability of PMMS on both surfaces at lower temperatures as revealed in **Figure 1(b)**. In both cases, the mass exchange between interfacial and core molecules might be affected in different way due to different adsorption-desorption processes occurring at the interface.

## IV. Conclusions

In conclusion, the behavior of PMMS infiltrated within various porous membranes was investigated as a function of pore size and wettability. Interestingly, it was found that as temperature decreases the contact angle of the examined oligomer on silica or alumina surfaces increases or decreases, respectively. Moreover, we demonstrated that the confined-induced fluctuations in segmental dynamics have a finite life-time as this effect is time-dependent allowing to modify the dynamics of nanomaterials according to our desired purpose. Quite interesting data delivered annealing experiments which revealed that annealing $\tau_{anneal}$ do not depend on the degree of supercooling or heating. However, the activation barrier of the equilibration process increases with lowering pore diameter and is much different in alumina ($d = 10$ nm) and silica pores ($d = 8$ nm). This finding can be understood in view of the contact angle measurements which showed that upon cooling wettability of PMMS on alumina and silica increases ($\theta \sim 12º$) and decreases ($\theta \sim 28º$), respectively. As a consequence, any mass exchange between interfacial and core molecules as well as adsorption/desorption processes occurring at the interface must be affected. This observation suggests the major impact of the wettability on the equilibration processes occurring in the polymer confined in mesoporous templates.

## AUTHOR INFORMATION

*Corresponding author: kamil.kaminski@us.edu.pl

**Notes.** The authors declare no competing financial interests.

**ACKNOWLEDGMENT.** K.K. is thankful for financial support from the Polish National Science Centre within the OPUS project (Dec. no 2022/47/B/ST4/00236).

**Supporting Information**. The Supporting Information file is available free of charge at http://pubs.acs.org. and contains experimental details and additional figures including DSC thermograms of examined PMMS, the comparison of the loss spectra were characterized by the same $\tau_\alpha$, the comparison of the data fitting the Havriliak-Negami (HN) functions for dielectric spectra, the time evolution of dielectric loss spectra collected upon the annealing experiments of PMMS within AAO templates of $d = 10$ nm and silica membranes of $d = 8$ nm.

## References

(1) Tarnacka, M.; Madejczyk, O.; Kaminski, K.; Paluch, M. Time and Temperature as Key Parameters Controlling Dynamics and Properties of Spatially Restricted Polymers. *Macromolecules* **2017**, *50* (13), 5188–5193. https://doi.org/10.1021/acs.macromol.7b00616.

(2) Nieto Simavilla, D.; Panagopoulou, A.; Napolitano, S. Characterization of Adsorbed Polymer Layers: Preparation, Determination of the Adsorbed Amount and Investigation of the Kinetics of Irreversible Adsorption. *Macromol Chem Phys* **2018**, *219* (3), 1–6. https://doi.org/10.1002/macp.201700303.

(3) Talik, A.; Tarnacka, M.; Grudzka-Flak, I.; Maksym, P.; Geppert-Rybczynska, M.; Wolnica, K.; Kaminska, E.; Kaminski, K.; Paluch, M. The Role of Interfacial Energy and Specific Interactions on the Behavior of Poly(Propylene Glycol) Derivatives under 2D Confinement. *Macromolecules* **2018**, *51* (13), 4840–4852. https://doi.org/10.1021/acs.macromol.8b00658.

(4) Adrjanowicz, K.; Paluch, M. Discharge of the Nanopore Confinement Effect on the Glass Transition Dynamics via Viscous Flow. *Phys Rev Lett* **2019**, *122* (17), 176101. https://doi.org/10.1103/PhysRevLett.122.176101.

(5) Tarnacka, M.; Dulski, M.; Geppert-Rybczyńska, M.; Talik, A.; Kamińska, E.; Kamiński, K.; Paluch, M. Variation in the Molecular Dynamics of DGEBA Confined within AAO Templates above and below the Glass-Transition Temperature. *The Journal of Physical Chemistry C* **2018**, *122* (49), 28033–28044. https://doi.org/10.1021/acs.jpcc.8b07522.

(6) Tarnacka, M.; Kaminska, E.; Kaminski, K.; Michael Roland, C.; Paluch, M. Interplay between Core and Interfacial Mobility and Its Impact on the Measured Glass Transition: Dielectric and Calorimetric Studies. *The Journal of Physical Chemistry C* **2016**, *120* (13), 7373–7380. https://doi.org/10.1021/acs.jpcc.5b12745.

(7) Panagopoulou, A.; Napolitano, S. Irreversible Adsorption Governs the Equilibration of Thin Polymer Films. *Phys. Rev. Lett.* **2017**, *119* (9), 97801. https://doi.org/10.1103/PhysRevLett.119.097801.

(8) Alexandris, S.; Papadopoulos, P.; Sakellariou, G.; Steinhart, M.; Butt, H. J.; Floudas, G. Interfacial Energy and Glass Temperature of Polymers Confined to Nanoporous Alumina. *Macromolecules* **2016**, *49* (19), 7400–7414. https://doi.org/10.1021/acs.macromol.6b01484.

(9) Napolitano, S.; Wübbenhorst, M. The Lifetime of the Deviations from Bulk Behaviour in Polymers Confined at the Nanoscale. *Nature Communications* **2011**, *2* (1), 260. https://doi.org/10.1038/ncomms1259.

(10) Napolitano, S.; Sferrazza, M. How Irreversible Adsorption Affects Interfacial Properties of Polymers. *Adv Colloid Interface Sci* **2017**, *247*, 172–177. https://doi.org/10.1016/j.cis.2017.02.003.

(11) Perez-De-Eulate, N. G.; Sferrazza, M.; Cangialosi, D.; Napolitano, S. Irreversible Adsorption Erases the Free Surface Effect on the Tg of Supported Films of Poly(4-Tert-Butylstyrene). *ACS Macro Lett* **2017**, *6* (4), 354–358. https://doi.org/10.1021/acsmacrolett.7b00129.

(12) Alcoutlabi, M.; McKenna, G. B. Effects of Confinement on Material Behaviour at the Nanometre Size Scale. *Journal of Physics: Condensed Matter* **2005**, *17* (15), R461–R524. https://doi.org/10.1088/0953-8984/17/15/R01.

(13) Tarnacka, M.; Kaminski, K.; Mapesa, E. U.; Kaminska, E.; Paluch, M. Studies on the Temperature and Time Induced Variation in the Segmental and Chain Dynamics in Poly(Propylene Glycol) Confined at the Nanoscale. *Macromolecules* **2016**, *49* (17), 6678–6686. https://doi.org/10.1021/acs.macromol.6b01237.

(14) Chat, K.; Adrjanowicz, K. The Impact of the Molecular Weight on the Nonequilibrium Glass Transition Dynamics of Poly(Phenylmethyl Siloxane) in Cylindrical Nanopores.

*The Journal of Physical Chemistry C* **2020**, *124* (40), 22321–22330. https://doi.org/10.1021/acs.jpcc.0c07053.

(15) Kardasis, P.; Sakellariou, G.; Steinhart, M.; Floudas, G. Non-Equilibrium Effects of Polymer Dynamics under Nanometer Confinement: Effects of Architecture and Molar Mass. *Journal of Physical Chemistry B* **2022**, *126* (29), 5570–5581. https://doi.org/10.1021/acs.jpcb.2c03389.

(16) Chat, K.; Adrjanowicz, K. The Impact of the Molecular Weight on the Nonequilibrium Glass Transition Dynamics of Poly(Phenylmethyl Siloxane) in Cylindrical Nanopores. *Journal of Physical Chemistry C* **2020**, *124* (40), 22321–22330. https://doi.org/10.1021/acs.jpcc.0c07053.

(17) Tarnacka, M.; Talik, A.; Kamińska, E.; Geppert-Rybczyńska, M.; Kaminski, K.; Paluch, M. The Impact of Molecular Weight on the Behavior of Poly(Propylene Glycol) Derivatives Confined within Alumina Templates. *Macromolecules* **2019**, *52* (9), 3516–3529. https://doi.org/10.1021/acs.macromol.9b00209.

(18) Talik, A.; Tarnacka, M.; Geppert-Rybczyńska, M.; Hachuła, B.; Kaminski, K.; Paluch, M. Impact of Confinement on the Dynamics and H-Bonding Pattern in Low-Molecular Weight Poly(Propylene Glycols). *The Journal of Physical Chemistry C* **2020**, *124* (32), 17607–17621. https://doi.org/10.1021/acs.jpcc.0c04062.

(19) Tarnacka, M.; Kamińska, E.; Paluch, M.; Kamiński, K. New Insights from Nonequilibrium Kinetics Studies on Highly Polar S-Methoxy-PC Infiltrated into Pores. *The Journal of Physical Chemistry Letters* **2022**, *13* (44), 10464–10470. https://doi.org/10.1021/acs.jpclett.2c02672.

(20) Tarnacka, M.; Jurkiewicz, K.; Hachuła, B.; Wojnarowska, Z.; Wrzalik, R.; Bielas, R.; Talik, A.; Maksym, P.; Kaminski, K.; Paluch, M. Correlation between Locally Ordered (Hydrogen-Bonded) Nanodomains and Puzzling Dynamics of Polymethysiloxane Derivative. *Macromolecules* **2020**, *53* (22), 10225–10233. https://doi.org/10.1021/acs.macromol.0c01289.

(21) Kipnusu, W. K.; Kossack, W.; Iacob, C.; Jasiurkowska, M.; Rume Sangoro, J.; Kremer, F. Molecular Order and Dynamics of Tris(2-Ethylhexyl)Phosphate Confined in Uni-Directional Nanopores. *Zeitschrift für Physikalische Chemie* **2012**, *226* (7–8), 797–805. https://doi.org/10.1524/zpch.2012.0287.

(22) Iacob, C.; Sangoro, J. R.; Papadopoulos, P.; Schubert, T.; Naumov, S.; Valiullin, R.; Kärger, J.; Kremer, F. Charge Transport and Diffusion of Ionic Liquids in Nanoporous Silica Membranes. *Physical Chemistry Chemical Physics* **2010**, *12* (41), 13798. https://doi.org/10.1039/c004546b.

(23) Havriliak, S.; Negami, S. A Complex Plane Analysis of α-Dispersions in Some Polymer Systems. *Journal of Polymer Science Part C: Polymer Symposia* **2007**, *14* (1), 99–117. https://doi.org/10.1002/polc.5070140111.

(24) Feder-Kubis, J.; Geppert-Rybczyńska, M.; Musiał, M.; Talik, E.; Guzik, A. Exploring the Surface Activity of a Homologues Series of Functionalized Ionic Liquids with a Natural Chiral Substituent: (−)-Menthol in a Cation. *Colloids and Surfaces A: Physicochemical and Engineering Aspects* **2017**, *529*, 725–732. https://doi.org/10.1016/J.COLSURFA.2017.06.040.

(25) Tarnacka, M.; Geppert-Rybczyńska, M.; Dulski, M.; Grelska, J.; Jurkiewicz, K.; Grzybowska, K.; Kamiński, K.; Paluch, M. Local Structure and Molecular Dynamics of Highly Polar Propylene Carbonate Derivative Infiltrated within Alumina and Silica Porous Templates. *The Journal of Chemical Physics* **2021**, *154* (6), 64701. https://doi.org/10.1063/5.0040150.

(26) Jackson, C. L.; McKenna, G. B. The Glass Transition of Organic Liquids Confined to Small Pores. *Journal of Non-Crystalline Solids* **1991**, *131–133* (PART 1), 221–224. https://doi.org/10.1016/0022-3093(91)90305-P.

(27) Adrjanowicz, K.; Kolodziejczyk, K.; Kiprop Kipnusu, W.; Tarnacka, M.; Urandu Mapesa, E.; Kaminska, E.; Pawlus, S.; Kaminski, K.; Paluch, M. Decoupling between the Interfacial and Core Molecular Dynamics of Salol in 2D Confinement. *The Journal of Physical Chemistry C* **2015**, *119* (25), 14366–14374. https://doi.org/10.1021/acs.jpcc.5b01391.

(28) Li, L.; Zhou, D.; Huang, D.; Xue, G. Double Glass Transition Temperatures of Poly(Methyl Methacrylate) Confined in Alumina Nanotube Templates. *Macromolecules* **2013**, *47* (1), 297–303. https://doi.org/10.1021/ma4020017.

(29) Chen, L.; Zheng, K.; Tian, X.; Hu, K.; Wang, R.; Liu, C.; Li, Y.; Cui, P. Double Glass Transitions and Interfacial Immobilized Layer in In-Situ-Synthesized Poly(Vinyl

Alcohol)/Silica Nanocomposites. *Macromolecules* **2010**, *43* (2), 1076–1082. https://doi.org/10.1021/ma901267s.

(30) Li, Q.; L. Simon, S. Surface Chemistry Effects on the Reactivity and Properties of Nanoconfined Bisphenol M Dicyanate Ester in Controlled Pore Glass. *Macromolecules* **2009**, *42* (10), 3573–3579. https://doi.org/10.1021/ma802808v.

(31) Politidis, C.; Alexandris, S.; Sakellariou, G.; Steinhart, M.; Floudas, G. Dynamics of Entangled Cis-1,4-Polyisoprene Confined to Nanoporous Alumina. *Macromolecules* **2019**. https://doi.org/10.1021/acs.macromol.9b00523.

(32) Glor, E. C.; Angrand, G. V; Fakhraai, Z. Exploring the Broadening and the Existence of Two Glass Transitions Due to Competing Interfacial Effects in Thin, Supported Polymer Films. *The Journal of Chemical Physics* **2017**, *146* (20), 203330. https://doi.org/10.1063/1.4979944.

(33) Zimny, S.; Tarnacka, M.; Kamińska, E.; Wrzalik, R.; Adrjanowicz, K.; Paluch, M.; Kamiński, K. Studies on the Molecular Dynamics at High Pressures as a Key to Identify the Sub-Rouse Mode in PMMS. *Macromolecules* **2022**, *55* (13), 5581–5590. https://doi.org/10.1021/acs.macromol.2c00378.

(34) Arndt, M.; Stannarius, R.; Groothues, H.; Hempel, E.; Kremer, F. Length Scale of Cooperativity in the Dynamic Glass Transition. *Physical Review Letters* **1997**, *79* (11), 2077–2080. https://doi.org/10.1103/PhysRevLett.79.2077.

(35) Uhl, M.; Fischer, J. K. H.; Sippel, P.; Bunzen, H.; Lunkenheimer, P.; Volkmer, D.; Loidl, A. Glycerol Confined in Zeolitic Imidazolate Frameworks: The Temperature-Dependent Cooperativity Length Scale of Glassy Freezing. *The Journal of Chemical Physics* **2019**, *150* (2), 24504. https://doi.org/10.1063/1.5080334.

(36) Arndt, M.; Stannarius, R.; Gorbatschow, W.; Kremer, F. Dielectric Investigations of the Dynamic Glass Transition in Nanopores. *Physical Review E* **1996**, *54* (5), 5377–5390. https://doi.org/10.1103/PhysRevE.54.5377.

(37) Schönhals, A.; Stauga, R. Dielectric Normal Mode Relaxation of Poly(Propylene Glycol) Melts in Confining Geometries. *Journal of Non-Crystalline Solids* **1998**, *235–237*, 450–456. https://doi.org/10.1016/S0022-3093(98)00657-7.

(38) Schönhals, A.; Stauga, R. Broadband Dielectric Study of Anomalous Diffusion in a Poly(Propylene Glycol) Melt Confined to Nanopores. *The Journal of Chemical Physics* **1998**, *108* (12), 5130–5136. https://doi.org/10.1063/1.475918.

(39) Schönhals, A.; Goering, H.; Schick, C.; Frick, B.; Zorn, R. Glassy Dynamics of Polymers Confined to Nanoporous Glasses Revealed by Relaxational and Scattering Experiments. *European Physical Journal E* **2003**, *12* (1), 173–178. https://doi.org/10.1140/epje/i2003-10036-4.

(40) Kuon, N.; Milischuk, A. A.; Ladanyi, B. M.; Flenner, E. Self-Intermediate Scattering Function Analysis of Supercooled Water Confined in Hydrophilic Silica Nanopores. *The Journal of Chemical Physics* **2017**, *146* (21), 214501. https://doi.org/10.1063/1.4984764.

(41) Shi, G.; Guan, Y.; Liu, G.; J. Müller, A.; Wang, D. Segmental Dynamics Govern the Cold Crystallization of Poly(Lactic Acid) in Nanoporous Alumina. *Macromolecules* **2019**, *52* (18), 6904–6912. https://doi.org/10.1021/acs.macromol.9b00542.

(42) Housmans, C.; Sferrazza, M.; Napolitano, S. Kinetics of Irreversible Chain Adsorption. *Macromolecules* **2014**, *47* (10), 3390–3393. https://doi.org/10.1021/ma500506r.

Supporting Information

# Is there a relationship between wettability and rates of equilibration of the H-bonded oligomer PMMS under confinement?

Sara Zimny[1], Magdalena Tarnacka[2], Monika Geppert-Rybczyńska[1], Kamil Kamiński[2*]

[1] *Institute of Chemistry, University of Silesia in Katowice, Szkolna 9, 40-006 Katowice, Poland*

[2] *Institute of Physics, University of Silesia in Katowice, 75 Pułku Piechoty 1, 41-500 Chorzów, Poland*

*\*Corresponding author:* kamil.kaminski@us.edu.pl

## Table of content

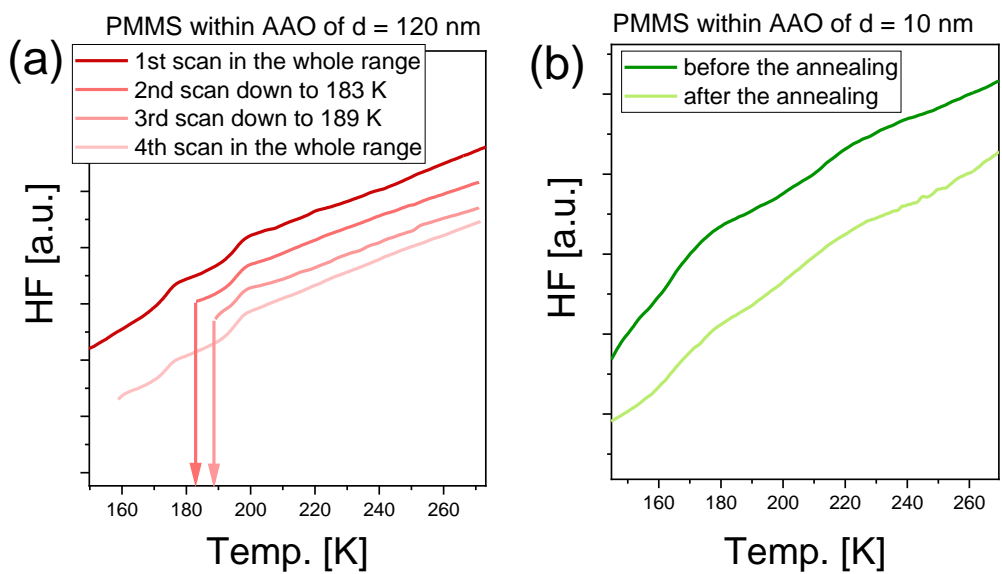

**Figure S1**. (a) DSC thermograms obtained for PMMS infiltrated within AAO templates of $d = 120$ nm. (b) DSC thermograms obtained for PMMS incorporated within AAO templates of $d = 10$ nm recorded before and after the annealing at $T_{anneal} \sim 240$ K ($T_{g,bulk} + 50$ K) for $t = 240$ min.

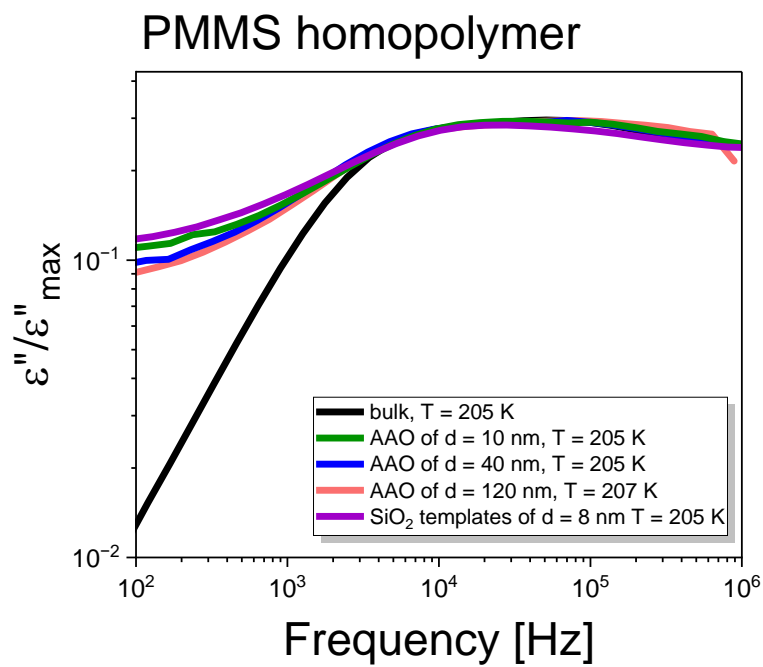

**Figure S2.** Comparison of the loss spectra obtained for examined PMMS infiltrated within AAO and silica templates of different pore size, $d$, at various temperatures conditions that were characterized by the same $\tau_\alpha$.

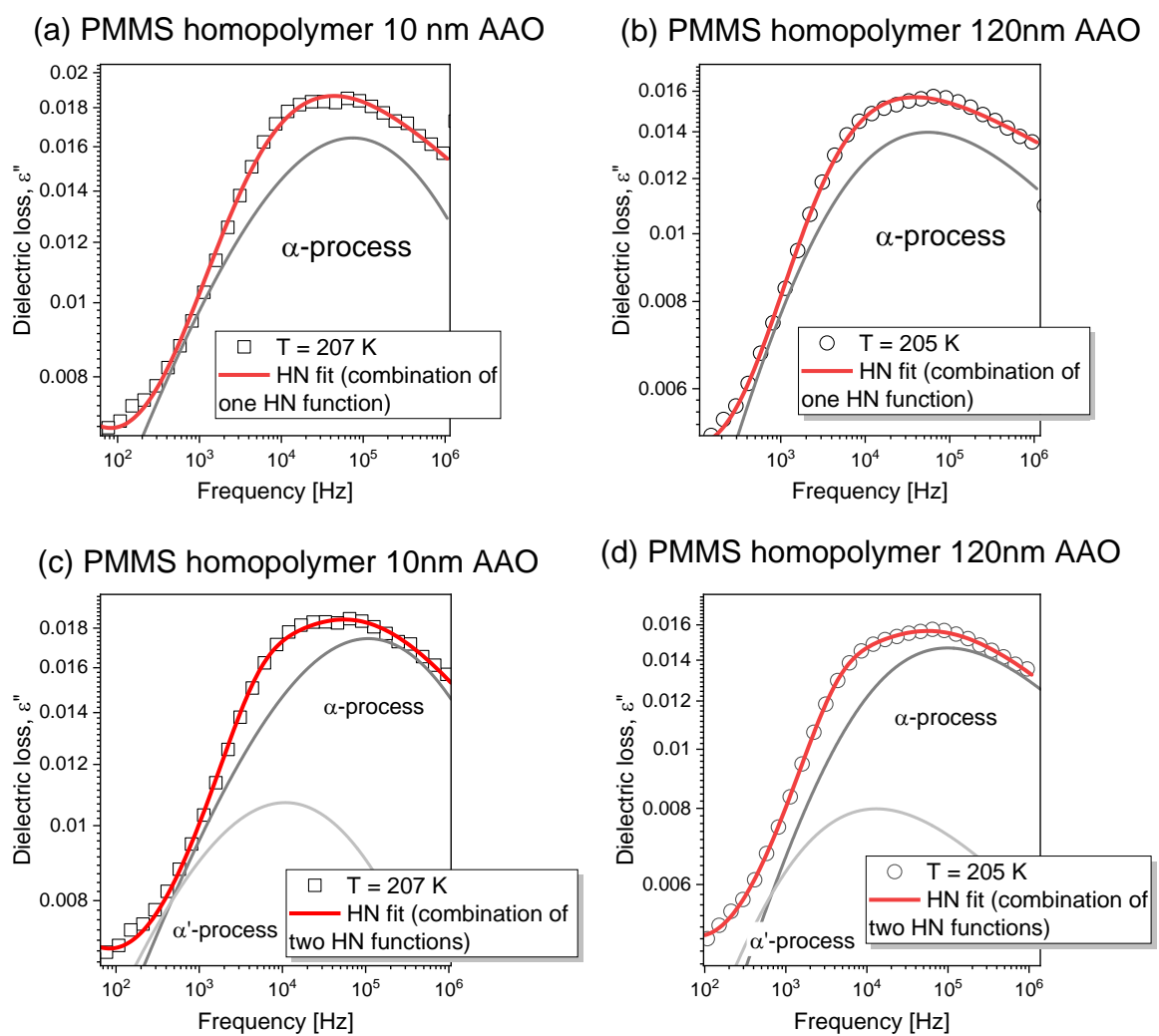

**Figure S3.** The comparison of the data fitting the use one (a, b) and two (c, d) Havriliak-Negami (HN) functions for dielectric loss spectrum of PMMS homopolymer confined within porous templates AAO ($d = 10, 120$ nm), collected at $T = 205$ K and $T = 207$ K.

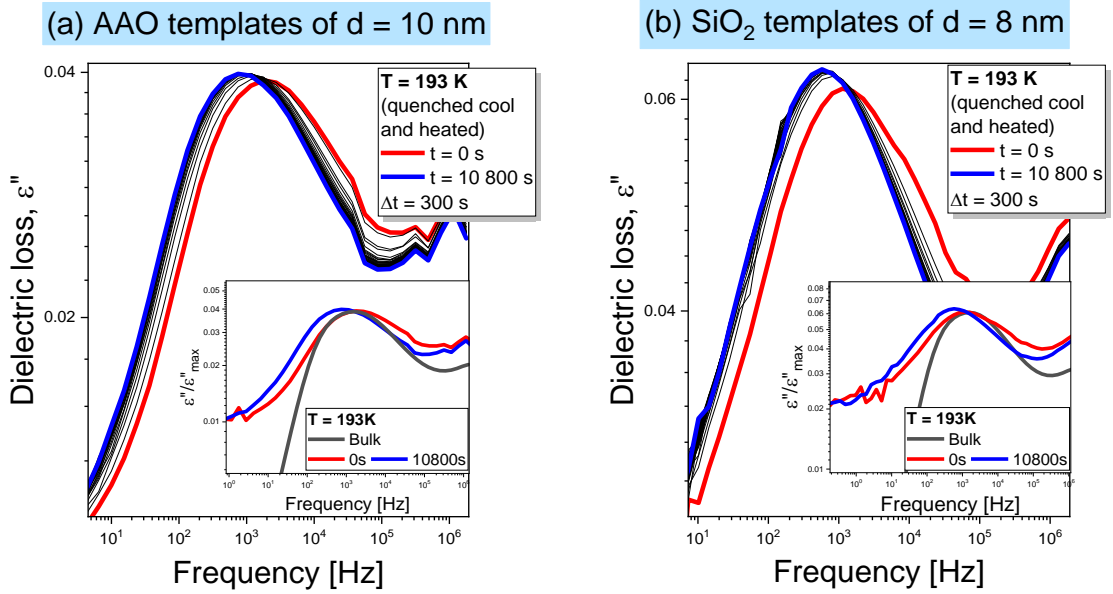

**Figure S4.** Time evolution of dielectric loss spectra collected upon the annealing experiments of PMMS within AAO templates of $d = 10$ nm and silica membranes of $d = 8$ nm at $T_{anneal} = 193$ K via different cooling rates. The insets show the comparison of dielectric loss spectra shapes before and after annealing in comparison to the bulk material.

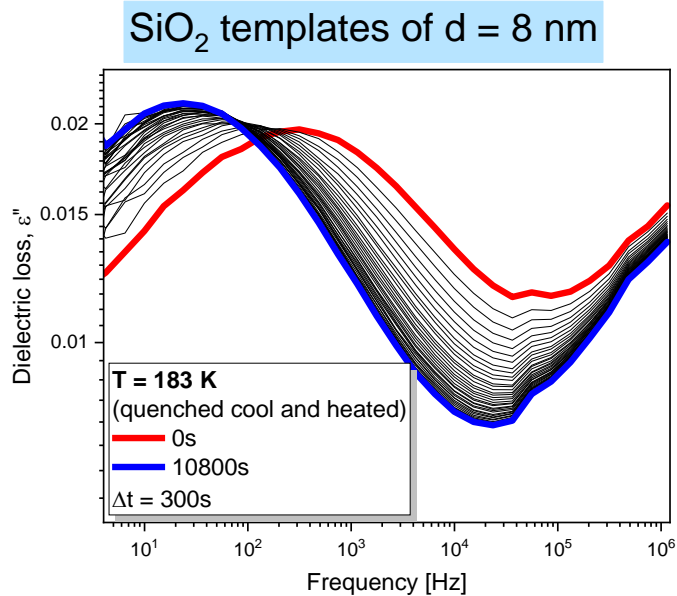

**Figure S5.** Time evolution of dielectric loss spectra collected upon the annealing experiments of PMMS within silica templates of $d = 8$ nm at $T_{anneal} = 183$ K via quenched cool and heated.